\documentclass[useAMS,usenatbib]{mn2e}

\usepackage{psfig}
\usepackage{graphicx}
\usepackage{times}
\usepackage{natbib}

\newif\ifAMStwofonts
\AMStwofontstrue

\newcommand{\be}{\begin{equation}}
\newcommand{\ee}{\end{equation}}
\newcommand{\ba}{\begin{eqnarray}}
\newcommand{\ea}{\end{eqnarray}}
\newcommand{\brr}{\begin{array}}
\newcommand{\err}{\end{array}}
\newcommand{\bc}{\begin{center}}
\newcommand{\ec}{\end{center}}
\newcommand{\hm}{\,h^{-1}{\rm Mpc}}
\newcommand{\hk}{\,h^{-1}{\rm kpc}}
\newcommand{\msun}{\,h^{-1}M_\odot}

\newcommand{\rvir}{\mbox{$R_{\rmn{vir}}$}}
\newcommand{\mvir}{\mbox{$M_{\rmn{vir}}$}}

\newcommand{\vel}{\,{\rm km\,s^{-1}}}

\newcommand{\Zfe}{\mbox{$Z_{\rmn{Fe}} \,$}}

\newcommand{\gadget}{{\footnotesize {\sc GADGET-2~}}}
\newcommand{\treesph}{{\footnotesize {\sc TreePM-SPH}}}
\newcommand{\Sal}{{\itshape Sal}}
\newcommand{\Kr}{{\itshape Kr}}
\newcommand{\AY}{{\itshape AY}}

\pssilent 

\newcommand{\mincir}{\raise
  -2.truept\hbox{\rlap{\hbox{$\sim$}}\raise5.truept \hbox{$<$}\ }}
\newcommand{\magcir}{\raise
  -2.truept\hbox{\rlap{\hbox{$\sim$}}\raise5.truept \hbox{$>$}\ }}
\newcommand{\siml}{\raise
  -2.truept\hbox{\rlap{\hbox{$\sim$}}\raise5.truept \hbox{$<$}\ }}
\newcommand{\simg}{\raise
  -2.truept\hbox{\rlap{\hbox{$\sim$}}\raise5.truept \hbox{$>$}\ }}

\newcommand{\ArY}{Arimoto--Yoshii~}

\newcommand{\Chandra}{{\sl Chandra~}}

\title[Evolution of the ICM metal content with simulations] 
  {Evolution of the metal content of the intra-cluster medium with
    hydrodynamical simulations}
  \author[Fabjan et al.] {D. Fabjan$^{1,2,3}$, L. Tornatore$^{1,2,3}$,
  S. Borgani$^{1,2,3}$, A. Saro$^{1,3}$ \& K. Dolag$^{4}$
  \\~\\
  $^1$ Dipartimento di Astronomia dell'Universit\`a di Trieste, via
  Tiepolo 11, I-34131 Trieste, Italy (fabjan,borgani,tornatore,saro@oats.inaf.it)\\
  $^2$ INAF -- Istituto Nazionale di Astrofisica, via Tiepolo 11,
  I-34131 Trieste, Italy\\
  $^3$ INFN -- Istituto Nazionale di Fisica Nucleare, Trieste, Italy\\
  $^4$ Max-Planck-Institut f\"ur Astrophysik, Karl-Schwarzschild Strasse
  1, Garching bei M\"unchen, Germany (kdolag@mpa-garching.mpg.de)\\
}

\begin{document}
\label{firstpage}
\maketitle

\begin{abstract}
  We present a comparison between simulation results and X--ray
  observational data on the evolution of the metallicity of the
  intra--cluster medium (ICM). The simulations of galaxy clusters have
  been carried out using a version of the \treesph\ \gadget code that
  includes a detailed model of chemical evolution, by assuming three
  different shapes for the stellar initial mass function
  (IMF). Besides the Salpeter (1955) IMF, we used also the IMF
  proposed by Kroupa (2001) and the top--heavier IMF by Arimoto \&
  Yoshii (1987).  We find that simulations predict significant radial
  gradients of the Iron abundance, \Zfe, which extend over the whole
  cluster virialized region. Using the Salpeter IMF, the profiles of
  \Zfe\ have an amplitude which is in a reasonable agreement with
  \Chandra observations within $0.2R_{500}$. At larger radii, we do
  not detect any flattening of the metallicity profiles.
  
  As for the evolution of the ICM metal abundance out to $z=1$, it
  turns out that the results based on the Salpeter IMF agree with
  observations. We find that the evolution of \Zfe\ in simulations is
  determined by the combined action of {\em (i) } the sinking of
  already enriched gas, {\em (ii)} the ongoing metal production in
  galaxies and {\em (iii)} the locking of ICM metals in newborn
  stars. As a result, rather than suppressing the metallicity
  evolution, stopping star formation at $z=1$ has the effect of
  producing an even too fast evolution of the emission--weighted ICM
  metallicity, with too high values of \Zfe\ at low redshift within
  $0.2R_{200}$. Finally, we compare simulations with the observed rate
  of type-Ia supernovae per unit B--band luminosity (SnU$_B$). We find
  that our simulated clusters do not reproduce the decreasing trend of
  SnU$_B$ at low redshift, unless star formation is truncated at $z=1$.
\end{abstract}


\section{Introduction}

The high quality X--ray observations of galaxy clusters from the
current generation of X--ray satellites are allowing now to trace in
detail the pattern of the metal enrichment of the intra--cluster
medium of galaxies \citep[e.g.,][for reviews]{2004cgpc.symp..123M,
2008arXiv0801.1052W}. In
turn, this information is inextricably linked to the history of
formation and evolution of the galaxy population as observed in the
optical/near--IR band \citep[e.g.,][and references
therein]{2004cgpc.symp..260R}. A number of independent observations
have established that significant radial gradients of the Iron
abundance are present in the central regions, $R\mincir 0.1 R_{500}$,
of relaxed clusters and groups \citep[e.g.,][]{2004A&A...419....7D,
  2005ApJ...628..655V, 2007MNRAS.380.1554R}, with enhancement of the
metallicity associated to the brightest cluster galaxies (BCGs), while
no evidence has been found that these gradients extend at larger
cluster--centric distances
\citep[e.g.,][]{2008A&A...478..615S}. Furthermore, deep exposures with
the Chandra and XMM--Newton satellites have now opened the possibility
of tracing the evolution of the ICM metal content within the central
regions out to the largest redshifts, $z\simeq 1.3$, where clusters
have been identified so far. \cite{2007A&A...462..429B} and
\cite{2008ApJS..174..117M} have analysed fairly large samples of
distant clusters, extracted from the Chandra archive and found that
the metallicity of the ICM within the central cluster regions has
increased by about 50 per cent since $z\simeq 1$.

This positive evolution of the ICM metallicity in the central cluster
regions is apparently in contradiction with the lack of significant
star formation at low redshift \citep[e.g.,][]{2006ApJ...652..216R}.
Based on a phenomenological approach, \cite{2005MNRAS.362..110E}
showed that the evolution of the ICM metallicity is in line with the
expectations from the observed cosmic rates of supernova (Sn)
explosions and of star formation.
\cite{2006ApJ...648..230L} combined observations of the evolution of
the ICM metallicity with data on the Sn rates and star formation rates
to infer the relative role played by type Ia and II Sn (SnIa and SnII
hereafter).

A different approach was pursued by other authors, which considered
gas--dynamical mechanisms that at relatively low redshift are
responsible for redistributing previously produced metals. For
instance, \cite{2008arXiv0802.0975C} suggested that clumps of
low--entropy highly enriched gas may sink in the central cluster
regions, thereby leading to an increase of the observed
emission--weighted metallicity. For instance, ram--pressure stripping
of the interstellar medium (ISM) of merging galaxies has been
suggested as a mechanism to pollute at relatively low redshift a
metal--poor ICM with highly enriched gas \citep[e.g.,][and references
therein]{2006A&A...452..795D}, while causing a morphological
transformation of cluster galaxies
\citep[e.g.,][]{2007MNRAS.tmpL..43C,2007MNRAS.380.1399R}.  Although
possible evidences of ram--pressure stripping of cluster galaxies have
been detected \citep[e.g.,][]{2007ApJ...659L.115C} the question
remains as to whether this mechanism dominates the evolution of the
ICM enrichment. Indeed, since ram pressure is expected to be more
efficient in high--temperature clusters, one expects an increasing
trend of metallicity with ICM temperature
\citep[e.g.,][]{1997ApJ...488...35R}. If any, observations suggest
that hotter systems have a relatively lower metallicity
\citep[e.g.,][]{2005ApJ...620..680B}, thus suggesting that
ram--pressure stripping is not the dominant process in enriching the
ICM.

It is clear that understanding the history of the ICM enrichment in
cosmological context, during the cluster hierarchical build up,
requires describing in detail the gasdynamics related to the merging
processes, while including a self--consistent treatment of star
formation and chemical evolution. In this context, cosmological
hydrodynamical simulations offer a unique means to capture in full
detail the complexity of these processes (e.g.,
\citealt{2003MNRAS.339.1117V,2004MNRAS.349L..19T,2006MNRAS.371..548R,2007MNRAS.382.1050T},
see \citealt{2008arXiv0801.1062B}, for a recent review). In their most
advanced versions, chemo--dynamical simulation codes treat the
production of different metal species, released by different stellar
populations by resorting to detailed stellar yields, also accounting
for the mass--dependent stellar lifetimes.

In this paper we will present results on the ICM metal abundance from
cosmological simulations of galaxy clusters, using the
chemo--dynamical version of the \gadget\ code
\citep{SP01.1,2005MNRAS.364.1105S}, which has been recently presented
by \cite{2007MNRAS.382.1050T} (T07 hereafter). We will compare the
simulations with observational results on the Iron abundance
profiles, \Zfe, of nearby clusters, on the evolution of the ICM
metallicity and on the SnIa rates. This comparison will be performed
with the aim of shading light on the relative role played by star
formation, feedback processes and gas dynamics in determining the
cosmic history of metal enrichment.

The plan of the paper is as follows. In Section 2 we review our
implementation of chemical evolution in the \gadget\ code and present
the main characteristics of the cluster simulations. Section 3 will be
devoted to the comparison between simulation results and
observations. After comparing the profiles of the Iron abundance, we
will concentrate on the evolution of the ICM metallicity. We will then
compare observations and simulation predictions on the rate of
SnIa. We will draw our conclusions in Section 4.

All values of Iron abundance that we will quote in the following are
scaled to the solar abundance value by \cite{1998SSRv...85..161G}.

\section{The simulations}

In this letter we present a set of simulations of four massive
isolated clusters, which have been identified in a Dark--Matter only
simulation having a box size $479 \hm$ \citep{2001MNRAS.328..669Y},
performed for a flat $\Lambda$CDM cosmological model with $\Omega_m =
0.3$, $h_{100} =0.7$, $\sigma_8 = 0.9$ and $\Omega_b = 0.04$.  The
four extracted Lagrangian regions, centred on these clusters with
virial masses\footnote{We define the virial mass $M_{\rm vir}$ as the
  mass contained within the virial radius \rvir. This is defined as
  the radius within which the average density $\rho_{\rm vir}$ is that
  predicted by the spherical collapse model (for the cosmology assumed
  in our simulations, $\rho_{\rm vir}\simeq 100 \rho_c$, with $\rho_c$
  the critical cosmic matter density). More in general, we define
  R$_{\Delta}$ to be the radius at which the density is $\Delta$ times
  the critical density $\rho_c$. Physical quantities with subscript
  $\Delta$ are computed within $R_{\Delta}$.} in the range $M_{\rm
  vir}=$1.0--2.3$\,\times 10^{15} \msun$, have been resimulated using
the Zoomed Initial Condition (ZIC) technique by \cite{TO97.2}, which
allows one to increase force and mass resolution in the regions of
interest. The high--resolution DM particles have mass
$m_{DM}=1.13\times 10^9 \msun$, and the barionic particles have been
added with a mass $m_{gas}=1.7\times 10^8 \msun$ in order to reproduce
the assumed cosmic barionic fraction. The basic characteristics of the
simulated clusters are summarized in Table \ref{tab:simul}.

The simulations are performed using the hydrodynamical Tree-SPH code
\gadget \citep{2005MNRAS.364.1105S} with the implementation of
chemical enrichment by T07. The Plummer--equivalent softening length
for gravitational force is set to $\epsilon = 5 \hk$ in physical units
from $z=2$ to $z=0$, while at higher redshifts is $\epsilon = 15 \hk$
in comoving units. The simulations include heating from a uniform
time-dependent UV background \citep{1996ApJ...461...20H} and
metallicity--dependent radiative cooling based on the tables by
\cite{1993ApJS...88..253S} for an optically thin plasma. The process
of star formation (SF hereafter) is described by the sub--resolution
multiphase model by \cite{2003MNRAS.339..289S}, for which the density
threshold for the onset of SF is set to $n_H=0.1\,$cm$^{-3}$.

While the relevant features of the chemical evolution model are
described here below, we address the reader to T07 for a more detailed
description. Metals are produced by SnII, SnIa and intermediate and
low--mass stars (ILMS hereafter), with only SnIa and SnII providing
energy feedback. We assume SnII to arise from stars having mass above
$8M_\odot$. As for the SnIa, we assume their progenitors to be binary
systems, whose total mass lies in the range (3--16)$M_\odot$. Metals
and energy are released by stars of different mass by properly
accounting for mass--dependent lifetimes. In this work we assume the
lifetime function proposed by
\cite{1993ApJ...416...26Pb}. We adopt the metallicity--dependent
stellar yields by \cite{1995ApJS..101..181W} for SnII, the yields by
\cite{1997A&AS..123..305V} for the ILMS and by
\cite{2003NuPhA.718..139T} for SnIa. The version of the code used for
the simulations presented here allowed us to follow H, He, C, N, O,
Mg, Si and Fe. Once produced by a star particle, metals are then
spread to the surrounding gas particles by using the B-spline kernel
with weights computed over 64 neighbours and taken to be proportional
to the volume of each particle. T07 verified with detailed tests that
the final results on the pattern of chemical enrichment are rather
insensitive to the weighting scheme (kernel shape and number of
neighbours) used to spread metals.

Our simulations include the kinetic feedback model implemented by
\cite{2003MNRAS.339..289S}. According to this scheme, SnII explosions
trigger galactic winds, whose mass upload rate is assumed to be
proportional to the star formation rate, $\dot{M}_W =\eta
\dot{M}_{\star}$. Therefore, fixing the parameter $\eta$ and the wind
velocity $v_W$ amounts to fix the total energy carried by the
winds. Our choice of $\eta=3$ and $v_W = 500\vel$ corresponds to
assume, for the initial mass function (IMF) by
\cite{1955ApJ...121..161S}, with SnII releasing $10^{51}$ ergs each,
nearly unity efficiency in powering galactic outflows.

In our comparison with observational data, we will first explore the
effect of changing the IMF. We use the IMF by
\cite{1955ApJ...121..161S} and that by \cite{1987A&A...173...23A}, for
which the number $N$ of stars per unit mass interval is defined as
$\varphi(m) \propto {\rm d}N/ {\rm d} m \propto m^{-(1+x)}$, with
$x=1.35$ and $x=0.95$ respectively. Furthermore, we also use the
multi--slope IMF proposed by \cite{2001MNRAS.322..231K} with $x=-0.7,
0.3$ and $1.3$ respectively for $m \le 0.08 M_{\odot}$, $0.08 \le m <
0.5 M_{\odot}$ and $m \ge 0.5 M_{\odot}$. Simulations based on the
Salpeter IMF have been run for the four clusters, while only
simulations of the g51 halo have been carried out for the other two
choices of the IMF. In the following, we label the runs that use the
Salpeter, \ArY\ and Kroupa IMFs with \Sal, \AY\ and \Kr\ respectively.

An important parameter entering in the model of chemical evolution is
the fraction $A$ of stars, in the mass range 0.8--8$\,M_\odot$,
belonging to binary systems which explodes as SnIa in the
single--degenerate scenario \citep{1983A&A...118..217G,
  1986A&A...154..279M}. For our reference runs we will use $A=0.1$, as
suggested by \cite{1995A&A...304...11M} to reproduce the observed ICM
metallicity (see also \citealt{2004ApJ...604..579P}). As we shall
discuss in the following, the simulation with the \AY\ IMF tends to
overproduce Iron. In the attempt to overcome this problem, we also
carried out a run with the \AY\ IMF using also $A=0.05$.

Simulations of galaxy clusters, which include the scheme of feedback
adopted here, are already known to produce an excess of low--redshift
star formation, mostly associated with the BCG
\citep[e.g.,][]{2005MNRAS.361..983R,2006MNRAS.373..397S}. This recent
star formation is expected to significantly affect the history of the
ICM enrichment. From one hand, it should provide an excess of
recent metal production, thus possibly enhancing the enrichment at
small cluster-centric radii. On the other hand, a recent star
formation is also expected to lock back in the stellar phase a
significant amount of highly enriched gas, which has shorter cooling
time, thus leaving in the hot ICM only relatively metal--poorer
gas. In order to quantify the effect of recent star formation of the
ICM enrichment history, we have also simulated the \Sal\ version of
the g51 cluster by switching off radiative cooling and star formation
below $z=1$, considering both the case in which already formed stars
keep producing metals with the appropriate lifetimes (CS run) and
the case in which also the metal production is stopped at the same
redshift (CMS run). While this prescription of suppressing
low--redshift star formation and metal production is admittedly
oversimplified, it allows us to address the following questions: {\em
  (i)} to what extent the SF excess in simulations affects the
enrichment evolution of the ICM? {\em (ii)} which is the role of
gas-dynamical processes in redistributing at relatively low redshift
the metals that have been produced at earlier epochs?


\begin{table}
\begin{center}
\begin{tabular}{|lrrr|}
\hline
Cluster & \mvir &  \rvir & $T^{sl}_{vir}$ \\ 
\hline
g1  & 1.49 &  2.33 & 7.90\\
g8  & 2.24 &  2.67 & 9.47\\
g72 & 1.34 &  2.26 & 6.23\\
g51 & 1.30 &  2.23 & 7.19\\
\hline
\end{tabular}
\end{center}
\caption{Characteristics of the simulated clusters
  at $z=0$. Column 1: cluster name; column 2: virial mass 
  (units of $10^{15}\msun$); column 3: virial radius (units of $\hm$);
  column 4: spectroscopic--like temperature within \rvir (keV, see
  \protect\cite{2004MNRAS.354...10M}, for its definition).}
\label{tab:simul}
\end{table}


\section{Results and discussion}

\subsection{Metallicity profiles of nearby clusters}
\label{Sec:met_prof}

The radial profiles of the metal abundance provide a very important
record of the chemical enrichment process in galaxy clusters. Indeed,
they are determined by the distribution of cluster galaxies, where
most of the metals are produced, by the mechanisms responsible for
their transport and diffusion from the star--forming regions (i.e.,
galactic ejecta, ram--pressure and viscous stripping, etc.) and by
other gas--dynamical processes which redistribute them on larger
scales (e.g. turbulence and sinking of enriched low--entropy
gas). Here we compare the profiles of the Iron abundance of simulated
galaxy clusters at $z=0$ with the observational results from \Chandra
data of a sample of nearby relaxed clusters analysed by
\cite{2005ApJ...628..655V}.

\begin{figure*}
\hbox{
\psfig{figure=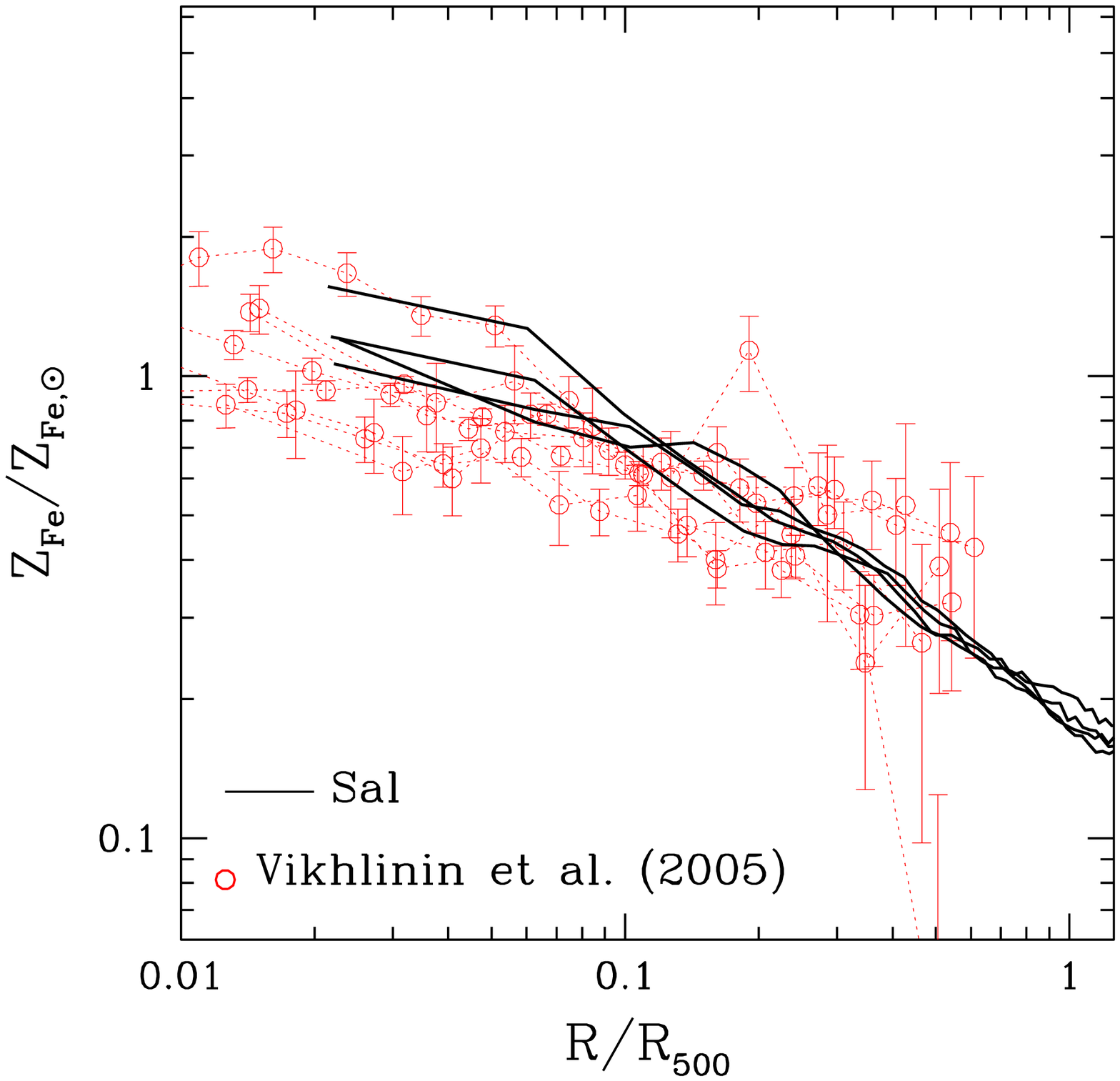,width=9.cm}
\psfig{figure=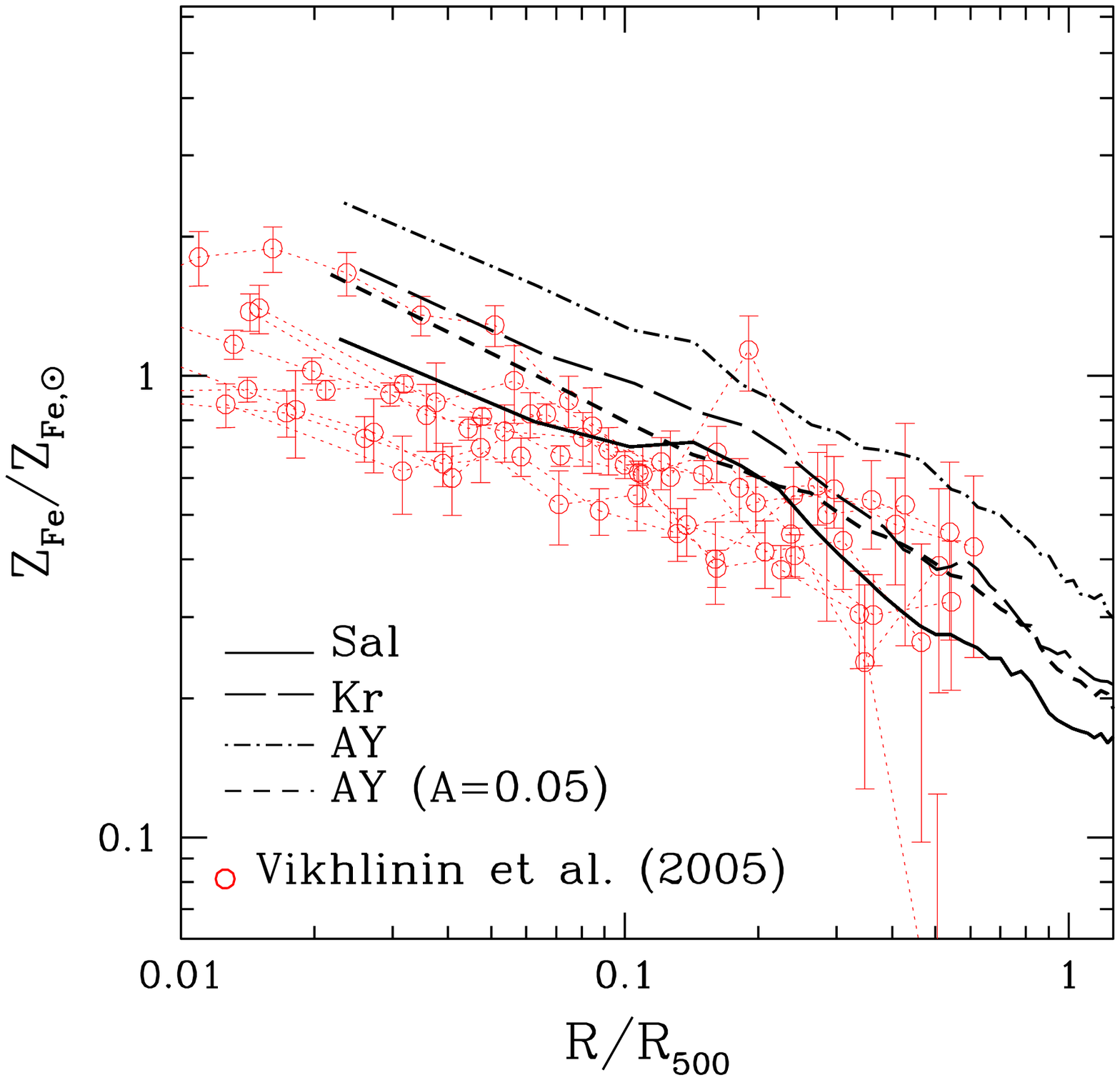,width=9.cm}
}
\caption{The Iron abundance profiles of simulated clusters at $z=0$
  compared to the Chandra observed profiles for 8 nearby clusters with
  $T>3$ keV \citep{2005ApJ...628..655V} (open circles with
  errorbars). In the left panel the solid curves correspond to the
  four clusters simulated using the \protect\cite{1955ApJ...121..161S}
  IMF. The right panel compares the results of the g51 cluster
  simulated using the IMF by \protect\cite{1955ApJ...121..161S}
  (solid), by \protect\cite{1987A&A...173...23A} with two different
  values for the fraction of binary stars ($A=0.1$: dot-dashed;
  $A=0.05$: short dashed) and by \protect\cite{2001MNRAS.322..231K}
  (long dashed).}
\label{Fig:prof_obs}
\end{figure*}

In Figure \ref{Fig:prof_obs} we compare the profiles of \Zfe\ from our
simulated clusters with the observed profiles of $8$ clusters having
temperature above $3$ keV \citep[see Table $1$
in][]{2005ApJ...628..655V}.  We point out that the analysis of
\cite{2005ApJ...628..655V} provided information on the total ICM
metallicity, i.e. without distinguishing the contribution from
different chemical species. However, at the typical temperatures of
these clusters and for the typical energy range where the spectral
analysis was performed (0.6--10 keV; see \citealt{2005ApJ...628..655V}), 
this observed metallicity is largely dominated by Iron. 
We want to stress the fact that the simulated clusters are 
dynamically relaxed (with the last major merger undergone before $z=0.5$) 
and therefore suitable for the comparison with this set of observed clusters.

As shown in the left panel of Fig.\ref{Fig:prof_obs}, the simulations
based on a \cite{1955ApJ...121..161S} IMF produce profiles which are
in reasonable agreement with observations. The scatter among the four
simulated clusters is quite small, with some increase in the central
regions, $R \, \mincir 0.1 \, R_{500}$. Although observations seem to
have a larger scatter, it is not clear how much observational
uncertainties contribute to it. Changing the IMF (right panel of
Fig. \ref{Fig:prof_obs}) clearly turns into a change of the overall
amount of the Iron abundance at all radii, with both the Kroupa and
the \AY\ IMFs producing too high profiles.  The larger amount of Iron
found for these two IMFs is due to the fact that, once
  normalized, they both predict a larger number of supernovae
  contributing to the Iron production, with respect to the Salpeter
  one.

Besides producing more SnII,
the \AY\ and \Kr\ IMFs also produce a larger number of SnIa, since there
is a significant overlap between the mass range relevant for SnIa and
the mass range where these two IMFs are higher than the Salpeter one.

As for the relative roles of SnIa, SnII and ILMS in the ICM
enrichment, we verified that SnIa contribute for about 70 per cent of
the Iron contained in the diffuse medium within $R_{500}$ for the
Salpeter IMF. This fraction decreases to about 65 per cent for
the Kroupa IMF and to about 55 per cent for the Arimoto--Yoshii
IMF. Since SnIa provide a major contribution to the Iron production,
our results are quite sensitive to the choice of the fraction $A$ of
stars in binary systems. As a matter of fact, this fraction can be
considered as a free parameter in a model of chemical evolution.
Following a phenomenological approach, for each choice of the IMF its
value is determined by the requirement of reproducing some
observational data. In our case, we note that decreasing $A$ from 0.1
to 0.05 induces a significant decrease of the \Zfe\ profile. This sort
of degeneracy between the IMF shape and the fraction of binary stars
can be broken by looking at the relative abundance of $\alpha$
elements with respect to Iron. For instance, since Oxygen is
essentially produced by SnII, we expect a top--heavier IMF to provide
values of O/Fe higher than for a top--lighter IMF. If we suppress the
number of SnIa for the top--heavier IMF, by decreasing the value of
$A$, we further increase the O/Fe ratio, thus allowing to distinguish
this case from that of a top--lighter IMF with a higher $A$. 
We deserve a forthcoming paper to a detailed comparison of simulations
with observations of relative abundances for nearby clusters.

The runs with the Salpeter IMF provide results in closer agreement
with the Chandra data, although in all cases the profiles of the
simulated clusters are somewhat steeper than the observed ones, with
negative gradients extending at least out $R_{500}$ and beyond. This
result is at variance with the recent claim by
\cite{2008A&A...478..615S} who found no evidence for the presence of
metallicity gradients at scales $\magcir 0.1R_{500}$ from the analysis
of a catalog of 70 clusters observed with XMM--Newton. In the next
section we will compare simulated and observed results on the
evolution of the ICM metallicity at small radii, where the simulated
and the observed metallicity gradients are in reasonable agreement. If
confirmed by independent analyses, the lack of abundance gradients at
relatively large radii will provide a non--trivial constraint for
chemo--dynamical models of the ICM enrichment.

Limited numerical resolution could lead to an underestimate of
high--redshift enrichment from a pristine population of relatively
small under--resolved galaxies. This high--$z$ enrichment should be
rather uniform and, therefore, should soften the metallicity
gradients. Indeed, T07 found that increasing resolution provides
progressively shallower metallicity profiles. However, the effect is
visible only at radii $\magcir 0.5 R_{500}$, while being negligible at
smaller radii, which are dominated by the star formation associated to
the BCG. Another possibility to soften metallicity profiles can be
provided by AGN feedback. For instance, \cite{2007arXiv0710.5574B}
analysed cosmological simulations of galaxy groups, which include the
effect of energy feedback from gas accretion onto black holes. They
found that the effect of this feedback is to redistribute the hot gas,
driving it from the inner regions, where it should be more enriched,
to the outer part of the halo, and to lower the star formation in the
inner region. \cite{2006MNRAS.366..397S} used a similar feedback
scheme, in which energy is used to trigger the formation of high
entropy bubbles. These bubbles rise buoyantly in the ICM, giving rise
to a redistribution of the central metal--enriched gas \citep[see also
][]{2007MNRAS.375...15R}. Clearly, in this case the request is that
the redistribution of metals should not be so efficient as to 
destroy the metallicity gradients in the central cluster regions
\cite[e.g.,][]{2004A&A...416L..21B}.

\begin{figure*}
\hbox{
\psfig{figure=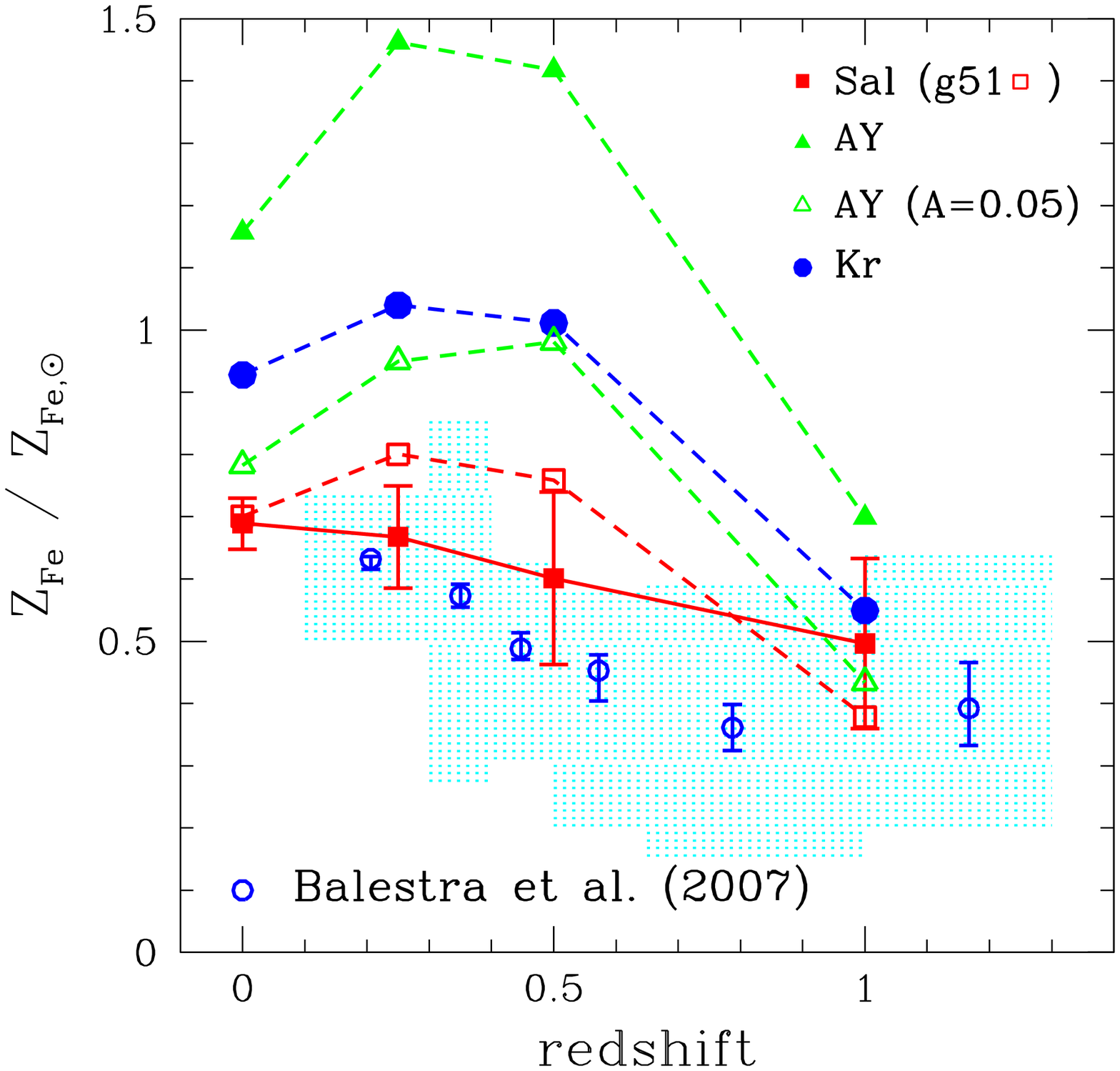,width=9.cm}
\psfig{figure=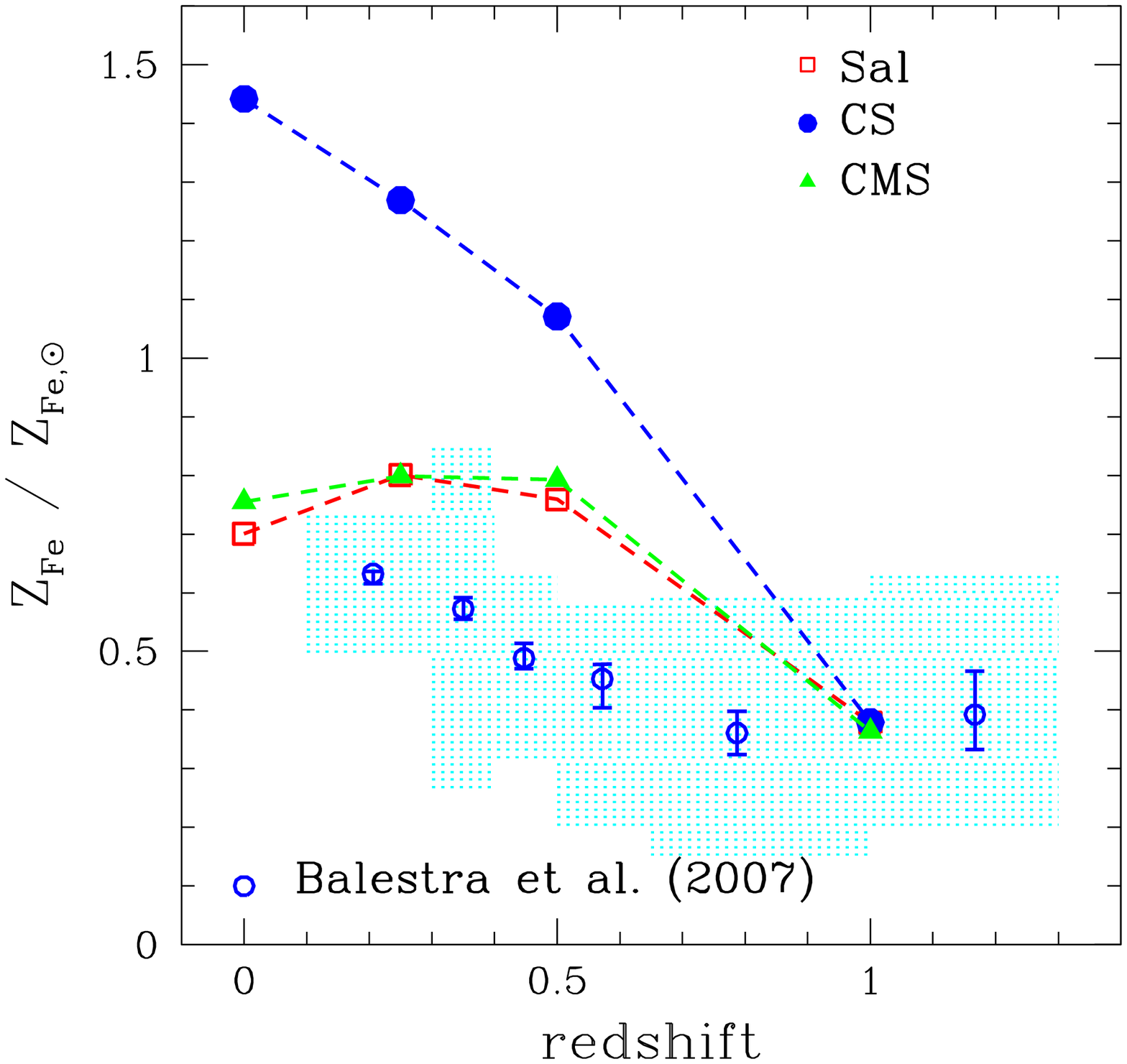,width=9.cm}
}
\caption{The comparison between observations and simulations for the
  evolution of the Iron abundance, \Zfe. Observational results from
  \protect\cite{2007A&A...462..429B} are shown with open circles, with
  errorbars corresponding to the $1\sigma$ uncertainty in the combined
  spectral fit performed for all the clusters falling within each
  redshift bin. The shaded area is the r.m.s. scatter among the
  measured metallicities within the same redshift intervals. Left
  panel: the dependence of the simulation results on the stellar
  IMF. The filled squares show the average over the simulated
  clusters, assuming a \protect\cite{1955ApJ...121..161S} IMF (\Sal),
  with errorbars indicating the r.m.s. scatter over the four
  objects. For the g51 cluster only, the open squares are for the run
  with \protect\cite{1955ApJ...121..161S} IMF, the filled and open
  triangles are for the \protect\cite{1987A&A...173...23A} IMF (\AY)
  with $A=0.1$ and $A=0.05$ for the binary fractions, respectively,
  while the filled circles are for the run with the
  \protect\cite{2001MNRAS.322..231K} IMF (\Kr). Right panel: the effect
  of stopping star formation and metal production at low redshift on
  the evolution of \Zfe\ for the g51 cluster, using the
  \protect\cite{1955ApJ...121..161S} IMF. Results for the reference (\Sal)
  run are shown with the open squares. The filled circles are for the
  simulation with radiative cooling and star formation stopped at
  $z=1$ (CS), while the filled triangles are for the run in which also
  the metal production is turned off at $z=1$ (CMS).}
\label{Fig:ZFe_red}
\end{figure*}

\subsection{Evolution of the ICM metallicity}
\label{Sec:evol_met}

In this Section we compare the simulation predictions on the evolution
of the ICM metallicity with the observational results by
\cite{2007A&A...462..429B}. These authors analysed \Chandra
observations of 56 clusters at $z>0.3$ (with the addition of
XMM--Newton observations for clusters at $z>1$) having temperatures
above $3$ keV. They measured the metallicity in the central regions,
with a typical extraction radius of $0.15$--$0.3\,R_{180}$, chosen
object-by-object so as to maximize the signal-to-noise ratio (see also
\citealt{2008ApJS..174..117M}, for a similar analysis). For the
low--redshift reference value, \cite{2007A&A...462..429B} combined
this set of distant clusters with a mix of cool-core and non cool-core
clusters at lower redshift. They also pointed out that the
  observed decrease of \Zfe with redshift is not induced by a decrease
  of the fraction of cool--core clusters in the past. Therefore, we
  expect that no significant bias is introduced when comparing the
  observed evolution with that traced by our set of relaxed simulated
  clusters. Since it is quite difficult to define a common extraction
radius for the observed clusters, we decided to adopt a value of $0.2
R_{180}$ in the analysis of the simulated clusters. We verified that
our conclusions are left unchanged by varying this radius in the range
0.15--0.3$\,R_{180}$.

In the left panel of Figure \ref{Fig:ZFe_red} we compare the
observational results by \cite{2007A&A...462..429B} with the
predictions of our simulations for different choices of the
IMF. Observations and simulations are compared here by using the
emission--weighted definition of metallicity, with emissivity of each
gas particle computed in the 0.5--10 keV energy band. In principle,
this comparison would require extracting synthetic spectra from the
simulated clusters and then measure the metallicity by fitting these
spectra to a single--temperature and single--metallicity plasma
models, as done in the analysis of observational data. An analysis of
this type has been recently presented by \cite{2007arXiv0707.2614R}
and showed that, at least for Iron, the emission--weighted estimator
gives results quite close (within about 10 per cent) to those obtained
from the spectral--fitting analysis.

Interestingly, we note in all simulations a significant increase of
metallicity in the cluster central regions below redshift unity. The
runs based on the \Sal\ IMF provide results close to observations. On
the contrary both the \Kr\ and the \AY\ IMFs predict too high
abundances for the g51 cluster at all redshifts, with a very strong
evolution at $z\magcir 0.5$ (we note that g51 is the cluster with the
highest metallicity at $z=0.25$ and 0.5, among the four simulated
objects). This higher abundance for the \AY\ and \Kr\ IMFs is in line
with the correspondingly higher profiles found at $z=0$ (see
Fig. \ref{Fig:prof_obs}). Again, decreasing the binary fraction in the
\AY\ run to $A=0.05$ causes a significant decrease of the Iron
abundance at all redshifts.

At lower redshifts, $z>0.5$, g51 shows a slower evolution, even
followed by an inversion below $z\simeq 0.25$. The reason for this
inversion lies in the quiet accretion history of this clusters below
$z=0.5$. Since no highly enriched gas clumps, associated to major
merger event, reached the central regions of g51 since $z=0.5$, the
only gas accreted there is characterized by a relatively low metal
abundance. For instance, in the \Sal\ run of g51 we verified that
beetween $z=0.25$ and $z=0$ about $6\times 10^{12}M_\odot$ of gas,
having an average metallicity value $Z_{Fe}\simeq 0.64\,Z_{Fe,\odot}$,
flowed out of $0.2R_{180}$. In the same redshift interval, about
$10^{13}M_\odot$ of gas was accreted, with an average metallicity
$Z_{Fe}\simeq 0.42\,Z_{Fe,\odot}$. Therefore, in the absence of large
metal--enriched clumps reaching the cluster centre, gas mixing leads
to the accretion of relatively metal--poorer gas, thus turning into a
decrease of \Zfe.

\begin{figure*}
\centerline{
\psfig{figure=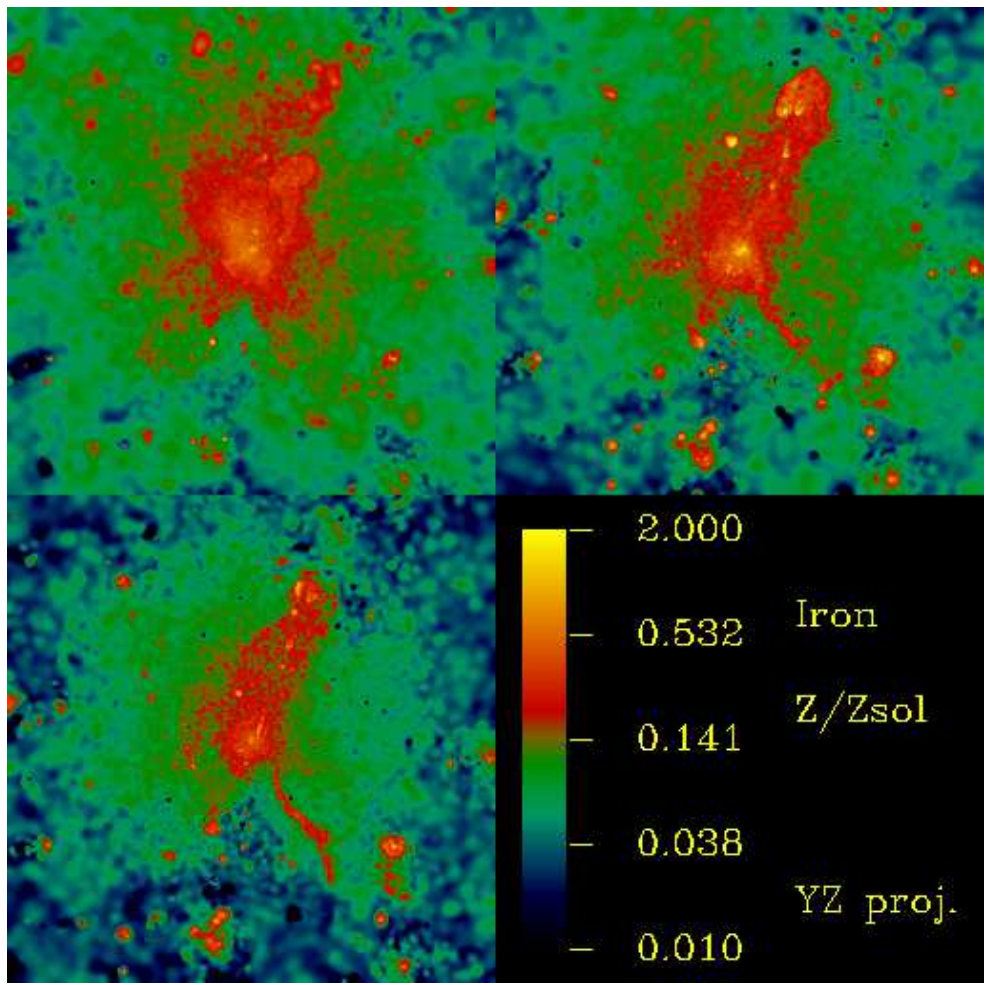,width=13.cm}
}
\caption{Projected maps of the emission--weighted Iron abundance for
  simulations of the g51 cluster, using the IMF by
  \protect\cite{1955ApJ...121..161S}, for the reference run (upper left), for
  the run with cooling and star formation stopped at $z=1$ (CS run;
  upper right) and for the run in which also the metal production is
  stopped at $z=1$ (CMS run; lower left). Each map is $2R_{\rm vir}$ a
  side.}
\label{Fig:maps}
\end{figure*}

Although the agreement between the runs based on the
\cite{1955ApJ...121..161S} IMF and observations is rather encouraging,
the question remains as to whether the positive evolution seen in the
simulations is just the spurious product of the excess of star
formation taking place in the central cluster regions. In order to
address this question, we compare in the right panel of
Fig.\ref{Fig:ZFe_red} the evolution of \Zfe for g51 when stopping star
formation and/or metal production at low redshift. Quite remarkably,
halting star formation below $z=1$ while allowing already formed stars
to keep releasing metals (CS run) has the effect of strongly
increasing the positive evolution of \Zfe in the central region of
g51, which turns out to be over--enriched by $z=0$. This leads to the
counter-intuitive conclusion that the lack of low--$z$ star formation
should generate an increase of the enrichment of the hot gas. In order
to investigate the origin of this increase, we show in Figure
\ref{Fig:maps} the emission--weighted metallicity maps of the
reference \Sal\ run of g51 at $z=0$, along with those of the CS and
CMS runs. Quite apparently, the metal distribution in the CS
simulation is more clumpy than in the reference run. At the cluster
centre, a high \Zfe\ is clearly visible, which boosts the central
emission weighted metallicity shown in Fig.\ref{Fig:ZFe_red}. Indeed,
while the emission--weighted \Zfe\ increases by about a factor two
within $0.2R_{180}$, we verified that the mass--weighted estimate
within the same radius increases only by about 10 per cent. In the
reference run, the metals released in the high density clumps
disappear from the hot diffuse medium due to the efficient gas
cooling. As a result, the reference run has a globally higher level
of diffuse enrichment, but a lower level of enrichment inside the
high--density gas clumps, which dominate the emission--weighted
estimate of \Zfe. These results demonstrate that the strongly
positive evolution of the emission--weighted metallicity in the CMS
run is driven by the accretion of highly enriched dense clumps.

Inhibiting also the production of metals below redshift unity (CMS
run) allows us to characterize the role played by gas--dynamical
processes in redistributing metals produced at higher redshift. As
shown in the bottom--left panel of Fig.\ref{Fig:maps}, metal clumps
are less pronounced than in the CS run. The global enrichment level of
the ICM is now significantly lower than in the reference run, although
an enhancement in the innermost regions is still visible. The
resulting mass--weighted metallicity within $0.2R_{180}$ at $z=0$
decreases by $\sim 60$ per cent with respect to the reference
run. Therefore, the stability of the emission--weighted metallicity is
due to the competing effects of a more clumpy distribution of metals
and of a decrease of the overall ICM metal budget.

The maps of Fig.\ref{Fig:maps} also illustrate the role of
gas--dynamical effects in redistributing highly enriched gas. Merging
clumps within the cluster virial region leave behind them
over--enriched tails of stripped gas, which is tempting to explain as
due to ram--pressure stripping. However, a significant contribution
could well be provided by viscous stripping. Since the SPH scheme is
known to be generally characterized by a large numerical viscosity,
this may induce an excess of gas stripping from merging
halos. \cite{2006MNRAS.371.1025S} showed that the effect of including
the Spitzer-Braginskii viscosity in the SPH, on the top of the
numerical viscosity, is indeed that of further increasing gas
stripping from merging halos. On the other hand,
\cite{2005MNRAS.364..753D} discussed an SPH scheme of reduced
viscosity. In this case, the increase of the ``turbulent'' stochastic
gas motions should provide a more efficient diffusion of metals from
star--forming regions \citep{2005MNRAS.359.1041R}, while making viscous
stripping less efficient. Although it is beyond the aim of this paper
to carry out an accurate analysis of the effect of viscosity on the
pattern of the ICM enrichment, there is no doubt that this aspect
deserve an accurate in-depth investigation.

\subsection{The SnIa rate}
\label{Sec:snia_rate}

The supernova rate represents a useful diagnostic to link the observed
evolution of the ICM metallicity to the past history of star formation
and to shed light on the relative contribution of SnIa and SnII in
releasing metals. In particular the ratio between the Sn rate and the
B-band luminosity, the so--called SnU$_B$, can be used to distinguish
the relative contribution of SnIa, which form in binary systems of
stars with masses in the range (0.8--8)$\,M_\odot$ and the short-living
massive stars that contribute substantially to 
the B--band luminosity of galaxies. In
this section we present a comparison between the results of our
simulated clusters and observational data of SnU$_B$ in galaxy
clusters from \cite{2002MNRAS.332...37G}, \cite{2008MNRAS.383.1121M}
and \cite{2007ApJ...660.1165S}.

\begin{figure*}
\hbox{
\psfig{figure=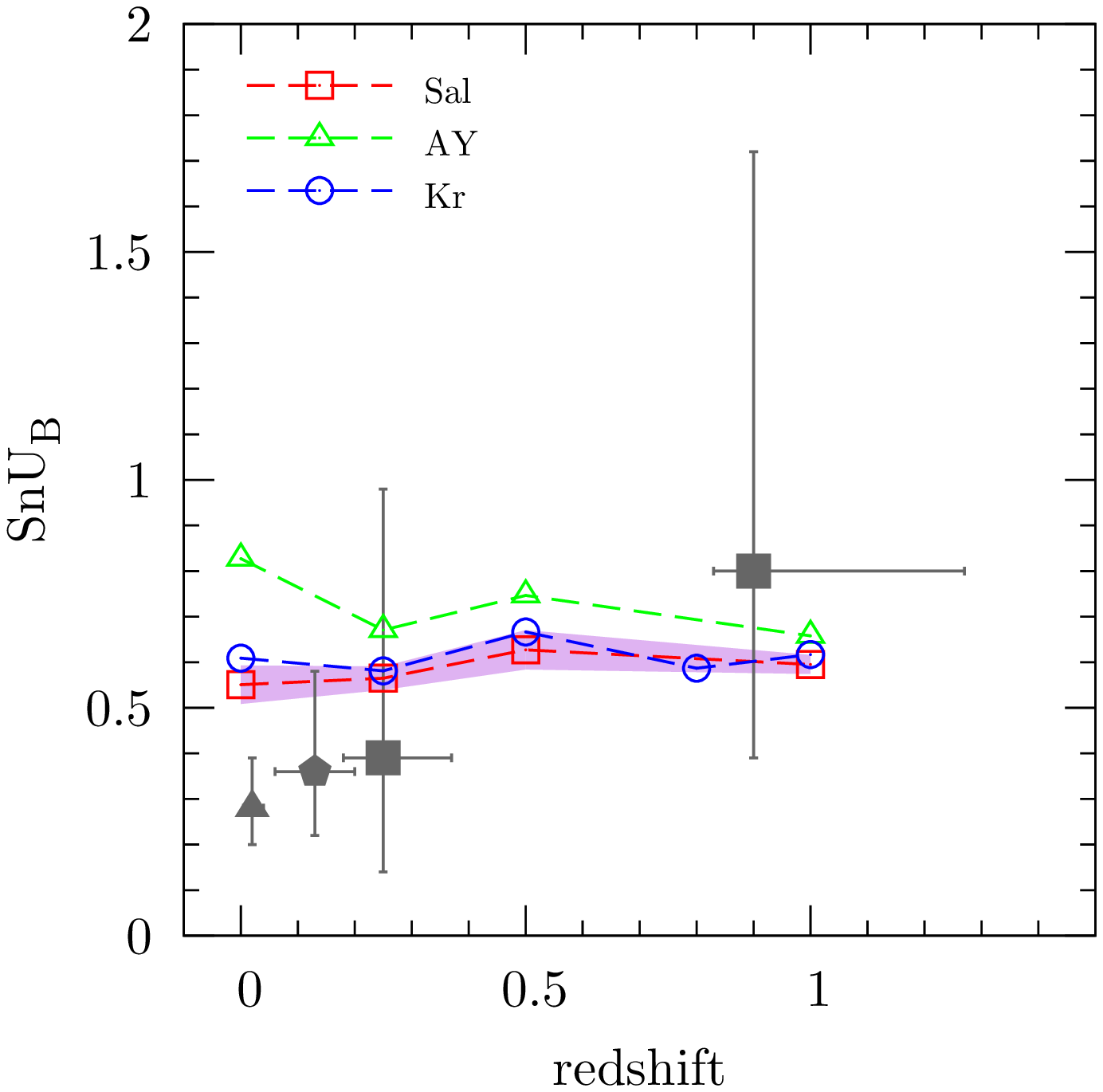,width=9cm}
\psfig{figure=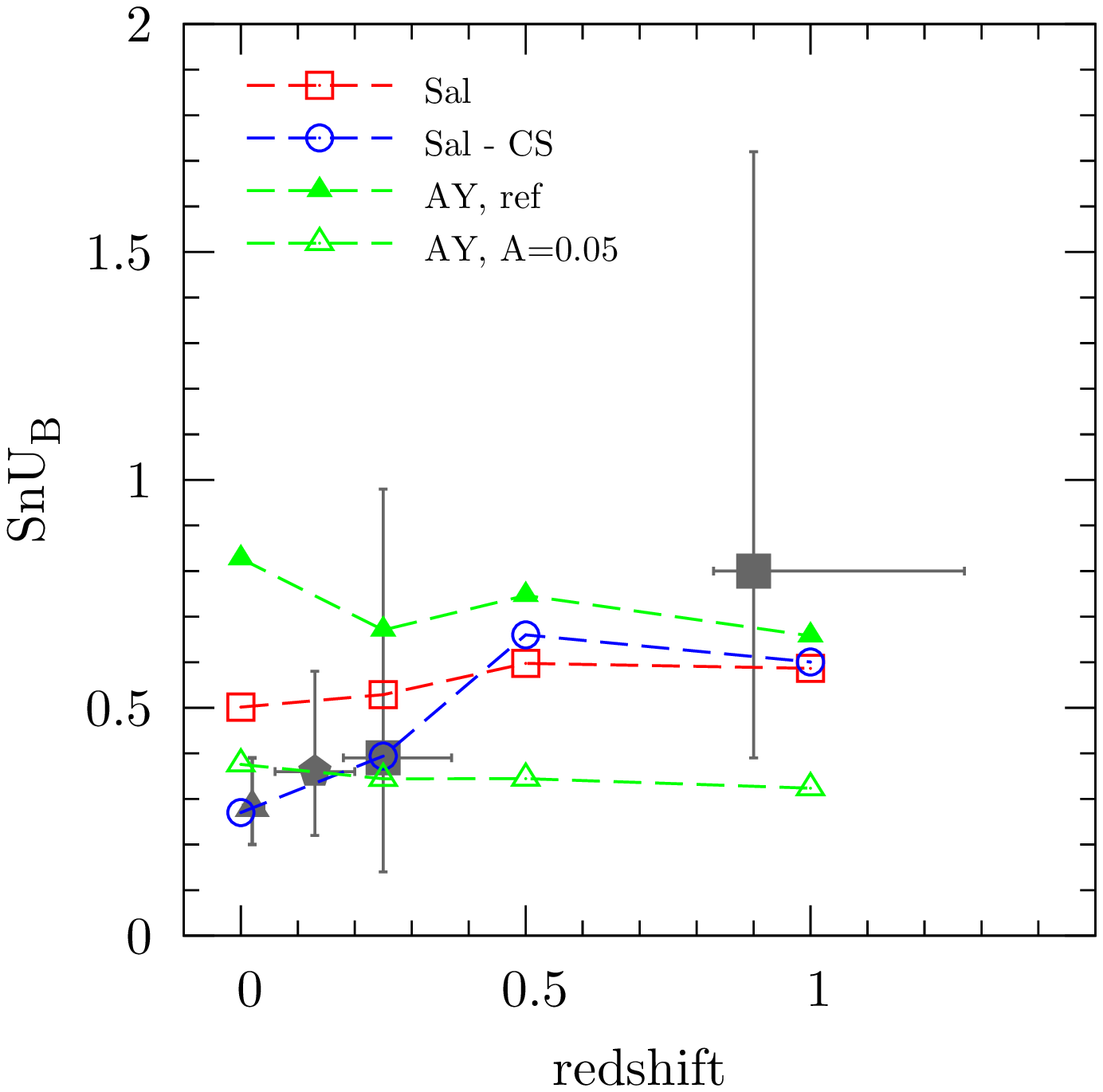,width=9cm}
}
\caption{Comparison between the observed and the simulated evolution
  of the SnIa rate per unit B--band luminosity (SnU$_B$). In both
  panels, filled symbols with errorbars refer to observational data
  from \protect\cite{2008MNRAS.383.1121M} (triangle),
  \protect\cite{2002MNRAS.332...37G} (squares) and
  \protect\cite{2007ApJ...660.1165S} (pentagon). Left panel: the
  effect of changing the IMF. The open squares are for the
  \protect\cite{1955ApJ...121..161S} IMF, the open triangles are for
  the top--heavy IMF by \protect\cite{1987A&A...173...23A} and the
  open circles for the IMF by \protect\cite{2001MNRAS.322..231K}. For
  the Salpeter IMF, the shaded area show the r.m.s. scatter evaluated
  over the four simulated clusters, while for the other two IMFs only
  the result for the g51 cluster is shown. Right panel: the effect of
  suppressing low--redshift star formation and of changing the binary
  fraction on the SnU$_B$ evolution of the g51 cluster. The open
  squares and the open circles are for the reference run with
  \protect\cite{1955ApJ...121..161S} IMF and for the same run with
  cooling and star formation stopped at $z=1$ (CS run),
  respectively. The filled and the open triangles are for the runs
  with \protect\cite{1987A&A...173...23A} IMF, using $A=0.1$ and
  $A=0.05$ for the fraction of binary stars, respectively.}
\label{Fig:SNU}
\end{figure*}

The simulation analysis finalized to compute the SnU$_B$ proceeds as
follows. For each star particle we know its formation redshift and
metallicity. Given the IMF and the lifetime function, this allows us
to compute the rate of SnIa exploding in each such
particle. Furthermore, using the spectrophotometric GALAXEV code
\citep{2003MNRAS.344.1000B} we also compute the B--band luminosity of
each star particle, which is treated as a Single Stellar Population
(SSP). Once SnIa rates and luminosities are computed for all the star
particles, we run the SKID substructure--finding algorithm
\citep{2001PhDT........21S} on their distribution to identify galaxies
as gravitationally bound groups of stars. All the star particles not
bound to galaxies take part of the intra-cluster diffuse stellar
component \citep[e.g.,][]{2007MNRAS.377....2M}.  We refer to
\cite{2006MNRAS.373..397S} for a detailed description of the procedure
to identify galaxies and assign broad--band luminosities to them. In
order to reproduce the observational procedure, we compute the SnU$_B$
values by also including the contribution of the SnIa arising from
diffuse stars, while the B--band luminosity is computed by including
only the contribution of the identified galaxies.

In the left panel of Figure \ref{Fig:SNU} we compare the SnU$_B$
values from the simulations with different IMFs with observational
data. In performing this comparison one potential ambiguity arises
from the definition of the extraction radius, within which
luminosities and SnIa rates are measured in observations, since
different authors use different aperture radii. To address this issue
we computed SnU$_B$ in the simulations
within $R_{\rm vir}$ and verified that the results are left
unchanged when using instead $R_{500}$. 

Observational data show a declining trend at low redshift. This is
generally interpreted as due to the quenching of recent star
formation, which causes the number of SnIa per unit B--band luminosity
to decrease after the typical lifetime of the SnIa progenitor has
elapsed.  On the other hand, our simulations predict a rather flat
evolution of the SnU$_B$, independently of the choice for the IMF,
which is the consequence of the excess of low--redshift star
formation. The runs based on the Salpeter and on the Kroupa IMF
produce very similar results. Although the Kroupa IMF produces a
higher rate of SnIa, due to its higher amplitude in the
(1--8)$\,M_\odot$ stellar mass range, this is compensated by the
higher values of $L_B$. These two IMFs both agree with the
observational data at $z \, \magcir 0.3$ within the large
observational uncertainties, while they overpredict the rates measured
for local clusters. Although the excess of recent star formation in
the central regions of our simulated clusters produces too blue BCGs
\citep{2006MNRAS.373..397S}, the large number of SnIa associated to
this star--formation overcompensate the excess of blue light. 

As for the simulation with the Arimoto--Yoshii IMF, it predicts an
even higher SnU$_B$ at low redshift. As shown in the right panel of
Fig.\ref{Fig:SNU}, decreasing the binary fraction to $A=0.05$
decreases the value of the SnU$_B$ by more than a factor 2. While this
helps in reconciling the simulation results with the low--redshift
data, it introduces a tension with the data at $z\sim 1$.

Truncating star formation at $z=1$ (right panel of
Fig.\ref{Fig:SNU}) has the desired effect of decreasing the value
of SnU$_B$ below $z=0.5$. Quite interesting, for $0.5\mincir z\mincir
1$ the decreasing trend of the SnIa rate is compensated by the
corresponding decrease of the B--band luminosity, while it is only at
$z>0.5$ that the decrease of the SnIa rate takes over causes the
decrease of the SnU$_B$ values.

\section{Conclusions}
We have presented results from cosmological SPH hydrodynamical
simulations of galaxy clusters with the purpose of characterising the
evolution of the chemical enrichment of the intra--cluster medium
(ICM) out to redshift $z\simeq 1$. The simulations have been performed
with a version of the \gadget\ code \citep{2005MNRAS.364.1105S}, which includes
a detailed model of chemical evolution \citep{2007MNRAS.382.1050T}. Our
simulations have been performed with the purpose of investigating
the effect of changing the chemical evolution model and the effect of
suppressing star--formation at $z<1$. The main results of our analysis
can be summarized as follows.

\begin{description}
\item[(a)] The Iron abundance profiles provided by simulations based
  on the \cite{1955ApJ...121..161S} IMF are in reasonable agreement
  with the results from Chandra observations of nearby clusters by
  \cite{2005ApJ...628..655V} at $R\mincir 0.2R_{500}$. Simulations
  based on the IMFs by \cite{2001MNRAS.322..231K} (\Kr) and
  \cite{1987A&A...173...23A} (\AY) both predict too high a
  normalization for these profiles. However, reducing the fraction of
  stars assumed to belong to binary systems suppresses the enrichment
  level, thus alleviating the disagreement of a top--heavy IMF with
  the observed \Zfe\ profiles. Our simulations always predict negative
  metallicity gradients extending out to $R_{500}$ and beyond,
  possibly in disagreement with XMM--Newton measurements of the Iron
  metal abundance at relatively large radii 
  \citep{2008A&A...478..615S}.
\item[(b)] All our simulations predict a positive evolution of the
  central Iron abundance, comparable to that observed by
  \cite{2007A&A...462..429B} (see also \citealt{2008ApJS..174..117M}). 
  Using a Salpeter IMF also provide an
  enrichment consistent with observations, while the Kroupa and
  Arimoto--Yoshii (\AY) IMFs overpredict the enrichment level at all
  redshifts. Again, this disagreement can be alleviated by
    decreasing the fraction of binary systems. It is worth reminding
    that the observed evolution of the Iron abundance is traced by
    using a mix of cool--core and non cool--core clusters, while our
    simulated clusters are all dynamically relaxed. Clearly, a fully
    self--consistent comparison would require simulating a
    representative population of clusters, having a variety of
    morphologies and dynamical states.
\item[(c)] Stopping cooling and star formation at $z=1$ (CS run) has
  the effect of producing a too strong positive evolution of the
  emission--weighted metallicity. Indeed, in the absence of star
  formation all the metals released at $z<1$ by long--living stars are
  no longer locked back in the stellar phase. As a result, metallicity is
  enhanced inside high--density halos and in the central cluster
  region. The clumpy metal distribution boosts the emission--weighted
  abundance estimate. This leads to the somewhat counter-intuitive
  conclusion that suppressing recent star formation has the effect of
  enhancing the positive evolution of the ICM metallicity.
\item[(d)] A comparison of the SnIa rate per unit B--band luminosity,
  SnU$_B$, show that our simulations are generally not able to
  reproduce the observed declining trend at low redshift. This result
  is explained by the excess of recent star formation taking place in
  the central regions of galaxy clusters. Indeed, excising star
  formation at $z<1$ produces an evolution of SnU$_B$ which is
  consistent with the observed one.
\end{description}

Cluster simulations which only include stellar feedback, like those
presented here, are well known to be at variance with a number of
observations, such as the temperature profiles in the cool core
regions and an large excess of recent star formation in the BCG.  Our
prescription to quench recent star formation is admittedly
oversimplified. A more realistic treatment would require introducing
energy feedback from gas accretion onto super-massive black holes,
which self--consistently follow the hierarchical build-up of the
cluster \citep[e.g.,][]{2006MNRAS.366..397S}. Still, our results
highlight that the positive evolution of the metal abundance in the
central regions of simulated clusters can not be simply interpreted as
a consequence of an excess of low--redshift star formation. In fact,
the evolution of the metallicity pattern is driven by the combined
action of the gas--dynamical processes, which redistribute already
enriched gas, and of star formation, which acts both as a source and
as a sink of metals. While hydrodynamical simulations probably provide
the most complete interpretative framework for observations of the
history of the ICM enrichment, they have still to improve in the
numerical accuracy for the description of relevant physical
processes. There is no doubt that the ever increasing supercomputing
power and efficiency of simulation codes should be paralleled by a
comparable advance in the reliability of the numerical description of
the relevant astrophysical and gas--dynamical processes. Only in this
way, simulations will become the standard tool to link ICM
observations to the global picture of cosmic structure formation.
 
\section*{Acknowledgments}
We are greatly indebted to Volker Springel for having provided us with
the non--public version of \gadget. We thank Silvia Ameglio for her
help in producing the metallicity maps of Fig.\ref{Fig:maps}, and
Alexey Vikhlinin for having provided the data points shown in
Fig.\ref{Fig:prof_obs}. We acknowledge useful discussions with Francesco 
Calura, Stefano Ettori, Alexis Finoguenov, Pasquale Mazzotta, 
Pierluigi Monaco, Elena Rasia and Paolo
Tozzi. The simulations have been carried out at the ``Centro
Interuniversitario del Nord-Est per il Calcolo Elettronico'' (CINECA,
Bologna), with CPU time assigned thanks to an INAF--CINECA grant and
to an agreement between CINECA and the University of Trieste, and on
the Linux Clusters at INAF in Catania and at the University of
Trieste. This work has been partially supported by the INFN PD-51
grant, by the INAF-PRIN06 Grant and by a ASI-AAE Theory grant.

\bibliographystyle{mn2e}
\bibliography{master}

\end{document}